\documentclass{tlp}
\usepackage{aopmath}

\newcommand{\sov}[1]{\overline{#1}}
\newcommand{\dov}[1]{\overline{\overline{#1}}}
\newcommand{\st}{\,|\;}

\newcommand{\calA}{\mathcal{A}}

\newcommand{\calF}{\mathcal{F}}
\newcommand{\calG}{\mathcal{G}}
\newcommand{\calC}{\mathcal{C}}

\newcommand{\vph}{\varphi}

\newcommand{\At}{\mathit{At}}

\newcommand{\bd}{\mathit{bd}}

\newcommand{\lla}{\leftarrow}

\newcommand{\n}{\mathit{not\;}}

\begin{document}

\submitted{1 January 2010}
\revised{1 January 2010}
\accepted{1 January 2010}

\long\def\comment#1{}

\title[Trichotomy and Dichotomy Results for Reasoning
with Logic Programs]{Trichotomy and Dichotomy Results on the Complexity of Reasoning
with Disjunctive Logic Programs}

\author[M. Truszczy\'nski]
{MIROS\L AW TRUSZCZY\'NSKI \\
Department of Computer Science\\
University of Kentucky\\
Lexington, KY~40506, USA\\
E-mail: mirek@cs.uky.edu
}

\pagerange{\pageref{firstpage}--\pageref{lastpage}}
\volume{\textbf{00} (00):}
\jdate{January 0000}
\setcounter{page}{1}
\pubyear{0000}

\maketitle

\label{firstpage}

\begin{abstract}
We present \emph{trichotomy} results characterizing the complexity of
reasoning
with disjunctive logic programs. To this end, we introduce a certain
definition schema for classes of programs based on a set of allowed
arities of rules. We show that each such class of programs has a finite
representation, and for each of the classes definable in the schema
we characterize the complexity of the existence of an answer set problem.
Next, we derive similar characterizations of the complexity of skeptical
and credulous reasoning with disjunctive logic programs. Such results
are of potential interest. On the one hand, they reveal some reasons
responsible for the hardness of computing answer sets. On the other hand,
they identify classes of problem instances, for which the problem is
``easy'' (in P) or ``easier than in general'' (in NP). We obtain similar
results for the complexity of reasoning with disjunctive programs under
the supported-model semantics.
\end{abstract}


\section{Introduction}

It is well known that the problem to decide whether a \emph{propositional}
disjunctive logic 
program has an answer set (the \emph{EAS problem}, for short) is 
$\Sigma_2^P$-complete \cite{eg95}. It is also well known that putting 
restrictions on the input instances may affect the complexity. For 
example, the EAS problem for normal logic programs is NP-complete
\cite{mt88}.

In this paper we study the complexity of the EAS problem for classes
of propositional disjunctive logic programs that can be defined by sets 
of program
rule \emph{arities}. We show that for each such class the problem is either 
in P, is NP-complete or is $\Sigma_2^P$-complete, and we fully characterize
the classes of programs that fall into each category.  We extend this 
result to establish similar characterizations for the problems of 
skeptical and credulous reasoning with disjunctive logic programs.
Such results are of potential interest. On the one hand, they reveal 
some reasons responsible for the hardness of computing answer sets; cf.
Lemmas \ref{lem1} and \ref{lem23}. On the other hand, they identify 
classes of problem instances, for which the problem is ``easy'' (in P)
or ``still easier than in general'' (in NP); cf. Lemmas \ref{lem21} and 
\ref{lem22}. 

We also consider the corresponding reasoning problems for 
the semantics of supported models \cite{mu92,bradix96jlp1,Inoue98-JLP} 
and obtain similar results. However,
the classification is simpler, as the reasoning problems under the 
semantics of supported models are confined to the first level of the
polynomial hierarchy. Thus, for each class of programs characterized
in terms of arities, the decision problems related to reasoning tasks 
for programs from that class with respect to supported models turn out
to be either in P or NP-complete (coNP-complete, depending on the task).

Our results can be regarded as \emph{trichotomy} (respectively, 
\emph{dichotomy}) results for the complexity of reasoning tasks in 
disjunctive logic programming. Similar
results are known for the complexity of reasoning in other 
formalisms: propositional satisfiability 
\cite{Schaefer78,BulatovJK05,CreignouKS01}, reasoning with minimal 
models \cite{Cadoli92}, default logic \cite{ChapdelaineHS07},
and abductive reasoning \cite{NordhZ08}. There is however, an important
distinction between those earlier papers and our approach. The results
contained there are concerned with the setting in which formulas are 
conjunctions of Boolean relations, and the set of models of a formula
is the intersection of the sets of models of its constituent relations
(in particular, it implies the monotonicity of inference from such 
formulas). The basic results concern the complexity of the satisfiability
problem for classes of formulas determined by sets of Boolean relations 
allowed as formula conjuncts. It turns out that there is a simple
characterization of all those classes, for which the problem is in P; 
moreover for all other classes the problem is NP-complete 
\cite{Schaefer78,BulatovJK05,CreignouKS01}. This result can be exploited
to characterize the complexity of reasoning with systems, in which basic
reasoning tasks reduce to series of satisfiability tests
\cite{Cadoli92,ChapdelaineHS07,NordhZ08}. In the setting of disjunctive
logic programs, these earlier results seem to be of little help. It is 
well known that the answer-set and the supported-model semantics are
\emph{nonmonotone} and so, 
logic programs under the answer-set semantics are \emph{not} conjunctions 
of their rules. Thus, it is unclear whether defining classes of programs in 
terms of semantic properties of individual rules could yield any useful
insights. 

Finally, we stress that we are concerned with propositional programs 
only. For results on the complexity of reasoning with disjunctive logic 
programs in the general (non-ground) case, we refer the reader to the
paper by Dantsin, Eiter, Gottlob, and Voronkov \citeyear{DantsinEGV01}.

\section{Preliminaries}

We fix an infinite countable set $\At$ of propositional variables. A
\emph{disjunctive program} (or simply, a program) over the set of atoms $\At$ is
a collection of \emph{disjunctive logic program rules}, that is, expressions 
of the form
\begin{equation}
\label{eq1}
r=a_1|\ldots|a_{k}\leftarrow b_1,\ldots,b_{m},\n c_1,\ldots,\n c_{n},
\end{equation}
where $a_i$, $b_i$ and $c_i$ are atoms from $\At$. The disjunction 
$a_1|\ldots|a_{k}$ is the \emph{head} of $r$ and the conjunction
$b_1,\ldots,b_{m},\n c_1,\ldots,\n c_{n}$ is the \emph{body} of $r$.
We call the triple $\alpha=[k,m,n]$ the \emph{arity} of $r$.
If $k\geq 1$, we call $r$ \emph{proper}. Otherwise, $k=0$ and $r$ is 
a \emph{constraint}. We allow the possibility that the head and body 
of a rule are both empty. Such rule (it is unique) is contradictory.

We recall that given a program $P$ and a set $M$ of atoms
(an interpretation), the \emph{reduct} of $P$ with respect to $M$, $P^M$,
is the program obtained by removing for all $c\in M$ all rules with a 
literal $\n c$ in the body and, then, removing all negative literals (negated 
atoms) from the bodies of all remaining rules. A set of atoms $M$ is an
\emph{answer set} of a disjunctive program $P$ if $M$ is a minimal model
of $P^M$ \cite{gl90b}. 

For a program $P$, we denote by $\sov{P}$ and $\dov{P}$ the programs 
consisting of all proper rules and of all constraints in $P$, respectively.
The following result is well known.

\begin{theorem}
\label{thm:c}
A set $M\subseteq\At$ is an answer set of a program $P$ if and
only if $M$ is an answer set of $\sov{P}$ and a model of $\dov{P}$.
\end{theorem}

One can define classes of logic programs by specifying arities of rules.
For instance, the set $\{[1,1,0],[0,1,0]\}$ defines the set of all Horn
programs with constraints such that each rule has at most one atom in 
the body. Some classes of programs do not have such a finitary 
representation in terms of arities. 
For instance, the class of all Horn programs with 
constraints can be defined by the set 
$\{[k,m,0]\st k\leq 1, 0\leq m\}$,
but there is no \emph{finite} set of arities that could be used instead.
To handle such cases, we introduce now a general representation schema 
for defining classes of programs in terms of sets of arities.

Let $U=\{0,1,\ldots\}\cup \{\infty\}$. We consider $U$ to be ordered by 
the relation $\leq$ (the standard $\leq$ ordering relation on non-negative
integers,
extended by $i\leq \infty$, for every $i=0,1\ldots$). Next, we define 
$\calA=\{[k,m,n]\st k,m,n\in U\}$. Thus, $\calA$ contains all
arities, as well as \emph{additional} triples --- those containing at 
least one occurrence of $\infty$. We refer to triples of that latter
sort as \emph{super\-arities}. We emphasize that super\-arities are 
\emph{not} arities as we do not consider infinitary rules. If $\alpha
\in\calA$, we write $\alpha_1$, $\alpha_2$ and $\alpha_3$ for the 
components of $\alpha$.

Let $\alpha,\beta\in\calA$. We define $\alpha\preceq\beta$ if
\begin{enumerate}
\item $\alpha_i \leq \beta_i$, for $i=1,2,3$, and 
\item if $\alpha_1=0$ then $\beta_1=0$.
\end{enumerate}
We write $\alpha\prec\beta$ when $\alpha\preceq\beta$ and
$\alpha\not=\beta$.

If $\Delta\subseteq\calA$, then we define $\calF(
\Delta)$ to be the set of all \emph{finite} programs $P$ that satisfy the 
following condition: for every rule $r\in P$ there is $\alpha\in\Delta$ 
such that $\alpha_r\preceq\alpha$, where $\alpha_r$ denotes the arity of $r$.
The condition (2) in the definition of
$\preceq$ allows us to distinguish between classes of proper programs
and classes of programs with constraints. Indeed, without the condition
(2), every class of programs of the form $\calF(\Delta)$ would contain 
constraints. With the condition (2), we can specify classes of
constraint-free programs by means of sets $\Delta$ such that for every 
$\alpha\in\Delta$, $\alpha_1\geq 1$. Including in $\Delta$ elements 
$\alpha$ with $\alpha_1=0$ yields classes of programs with constraints.
As there are classes of proper programs that are of interest (Horn 
programs and normal logic programs are typically defined as consisting 
of proper rules only), the distinction is needed and motivates the 
condition (2) in the definition of $\preceq$.

Using this schema we can define several important classes of programs.
For instance, the class of proper Horn programs can be described as
$\calF(\{[1,\infty,0]\})$ and the class of normal logic programs with
constraints as $\calF(\{[1,\infty,\infty],[0,\infty,\infty]\})$.

Our main goal in this paper is to determine the complexity of the EAS 
problem when input programs come from classes $\calF(\Delta)$, for 
$\Delta \subseteq\calA$. We also consider the corresponding problem 
for the case of supported models. 

\section{The Case of Finite $\Delta$}
\label{finite}

In this section we tackle the case when $\Delta$ is finite. We note that
given a finite set $\Delta\subseteq\calA$, the problem to decide the
membership of a program in the class $\calF(\Delta)$ is in P. 

We start by establishing upper bounds on the complexity of the 
EAS problem for classes $\calF(\Delta)$ given by some particular finite 
sets $\Delta$ of arities. Our first result is concerned with the 
following classes:
$\calF(\{[\infty,\infty,0],[0,0,0]\})$ --- the class of proper positive
disjunctive 
programs; $\calF(\{[1,\infty,0], [0,\infty,\infty]\})$ --- the class of 
programs whose every rule is either a proper Horn rule or a constraint;
$\calF(\{[\infty,1,0],$ $[0,1,0]\})$ --- the class of \emph{dual Horn 
programs}, that is, programs whose every rule, when viewed as a
propositional clause, is a dual Horn clause;
and $\calF(\{[i,j,0]\in \calA\st i+j\leq 2\})$ --- the class of positive 
programs whose every rule consists of at most two literals. For each of these classes of programs,
the EAS problem is easy (that is, in P).

\begin{lemma}
\label{lem21}
If $\Delta$ is one of:
\begin{enumerate}
\item $\{[\infty,\infty,0],[0,0,0]\}$ 
\item $\{[1,\infty,0], [0,\infty,\infty]\}$
\item $\{[\infty,1,0],[0,1,0]\}$
\item $\{[i,j,0]\in \calA\st i+j\leq 2\}$
\end{enumerate}
then the EAS problem for $\calF(\Delta)$ is in $\mathrm{P}$.
\end{lemma}
Proof: 
If $P\in\calF(\{[\infty,\infty,0],[0,0,0]\})$, then $P$ either contains
a contradictory rule and so does not have answer sets, or it is a proper 
positive program. In the latter case, $P$ has models and so, minimal 
models, too. It follows that the EAS problem is in P, in this case.

Next, let $P\in\calF(\{[1,\infty,0],[0,\infty,\infty]\})$. In this 
case, $\sov{P}$ is a proper Horn program. It has a least model, say 
$M$, which is the only answer set of $\sov{P}$. It is well known that
$M$ can be computed in polynomial time \cite{dg84}. Moreover, it can be
verified in polynomial time whether $M$ satisfies the constraint part
$\dov{P}$ of $P$. Thus, the assertion follows in this case, as well.

Thus, let us assume that
$P\in\calF(\{[\infty,1,0],[0,1,0]\})$ or $P\in
\calF(\{[i,j,0]\in\calA\st i+j\leq 2\})$. Then $P$ is a dual Horn program,
or $P$ is positive and every clause in $P$ consists of two literals. In 
each case, one can decide in polynomial time whether $P$ has a 
model.\footnote{For instance, for a dual Horn program $P$, the case that 
is perhaps less 
broadly known, one can compute in the bottom-up fashion the complement of 
its greatest model contained in $\At(P)$, or determine that no models exist 
(by means of a ``dual'' Dowling-Gallier algorithm).}
If the answer is 
``no,'' then $P$ has no answer sets. Otherwise, $P$ has a model, say $M$. 
Since $M$ is a model of $\sov{P}$, there is a subset $M'$ of $M$ such 
that $M'$ is a minimal model of $\sov{P}$. We have $\sov{P}^{M'}=\sov{P}$.
Thus, $M'$ is an answer set of $\sov{P}$. Since $M$ satisfies $\dov{P}$ 
and each rule in $\dov{P}$ is of the form $\lla a$ or $\lla a,b$, 
$M'$ satisfies $\dov{P}$, too. Thus, $M'$ is an answer set of $P$. Again,
the assertion follows. \hfill$\Box$

The second result establishes sufficient conditions for the EAS problem
to be in the class NP. It turns out to be the case for the following 
three classes of programs: $\calF(\{[1,\infty,\infty],
[0,\infty,\infty]\})$ --- the class of normal logic programs with constraints;
$\calF(\{[\infty,1,\infty],$ $[0,\infty,\infty]\})$ --- the class of programs 
whose reducts consist of proper dual Horn rules and constraints; and 
$\calF(\{[\infty,\infty,0],$ $[0,\infty, 0]\})$ --- the class of positive
programs.

\begin{lemma}
\label{lem22}
If $\Delta$ is one of:
\begin{enumerate}
\item $\{[1,\infty,\infty],[0,\infty,\infty]\}$
\item $\{[\infty,1,\infty],[0,\infty,\infty]\}$
\item $\{[\infty,\infty,$ $0],[0,\infty, 0]\}$
\end{enumerate}
then the EAS problem for $\calF(\Delta)$ is in $\mathrm{NP}$.
\end{lemma}
Proof: If $\Delta=\{[1,\infty,\infty],[0,\infty,\infty]\}$, 
$\calF(\Delta)$ consists of normal logic programs with constraints.
In this case, the result is well known \cite{mt88}.

Next, let $\Delta=\{[\infty,1,\infty],[0,\infty,\infty]\}$ and $P\in
\calF(\Delta)$. To prove the assertion it is enough to show that there
is a polynomial time algorithm for deciding whether a set of atoms $M
\subseteq\At(P)$ is an answer set of $P$. To this end, we note that $M
\subseteq\At(P)$
is a minimal model of $\sov{P}^M$ if and only if for every $a\in M$, the
program $\sov{P}^M\cup\{\;\lla a\}\cup\{\;\lla b \st b\in\At(P)\setminus M\}$
does not have a model. Since 
$\sov{P}^M\cup\{\;\lla a\}\cup\{\;\lla b \st b\in\At(P)\setminus M\}$ is
dual Horn, verifying whether $M$ is a minimal model of $\sov{P}^M$ can be 
accomplished in polynomial time. In 
addition, checking that $M$ is a model of $\dov{P}$ can be done in 
polynomial time, too. Thus, in this case, the assertion follows.

Finally, if $\Delta=\{[\infty,\infty,0],[0,\infty,0]\}$ and $P\in
\calF(\Delta)$, then deciding whether $P$ has an answer set is 
equivalent to deciding whether $P$ has a model. 
Indeed, if $M$ is an
answer set of $P$, then $M$ is a model of $P$. Conversely, if $M$ is a
model of $P$, then let $M'\subseteq M$ be a minimal model of $\sov{P}$
(such a model exists, as $M$ is a model of $\sov{P}$). Clearly, 
$\sov{P}^{M'}=\sov{P}$ and so, $M'$ is an answer set of $\sov{P}$.
Since $M' \subseteq M$ and $M$ is a model of $\dov{P}$, $M'$ is a model of 
$\dov{P}$ (it follows from the fact that every rule in $\dov{P}$ is of
the form $\;\lla a_1,\ldots,a_m$). Thus, $M'$ is an answer set of $P$.
Since the problem to decide whether $P$ has a model is in NP, the
assertion follows. \hfill$\Box$

While the upperbounds provided by Lemma \ref{lem22}(1) and (3) are not 
surprising (in fact, as we noted, Lemma \ref{lem22}(1) is well known),
Lemma \ref{lem22}(2) warrants additional comments. The class of programs
considered there, has not been identified before as one of the classes of
disjunctive programs for which the complexity of reasoning drops down to
the first level of the polynomial hierarchy. In the same time, it is an 
interesting class of programs. In particular, programs in this class are 
not, in general \emph{head-cycle free} \cite{bed94}.
 
Next, we will prove several lower-bound results. We will first exhibit 
classes of programs of the form $\calF(\Delta)$ for which the EAS problem
is NP-hard. To this end, we need a lemma establishing the NP-hardness of 
the SAT problem for some simple classes of CNF theories. For the most 
part, the result is folklore. We sketch an argument for the sake of 
completeness.
\begin{lemma}
\label{lem2}
The SAT problem restricted to each of the following classes of CNF theories
is NP-hard:
\begin{enumerate}
\item the class of all CNF formulas $\psi$ such that each
clause of $\psi$ is a disjunction of two negated atoms, or of at most three
atoms;
\item the class of all CNF formulas $\psi$ such that each
clause of $\psi$ consists of at most two negated atoms, or is a disjunction 
of two atoms, or is a disjunction of two atoms and one negated atom;
\item the class of all CNF formulas $\psi$ such that each
clause of $\psi$ consists of at most two atoms, or of one negated atom,
or is a disjunction of an atom and two negated atoms;
\item the class of all CNF formulas $\psi$ such that each
clause of $\psi$ is a disjunction of two atoms, or of at
most three negated atoms.
\end{enumerate}
\end{lemma}
Proof: We will only prove the case (3). The argument in
all other cases is similar.

Let $\vph$ be a CNF formula whose every clause has three literals, and
let $X$ be a set of atoms occurring in $\vph$. For each atom $z\in X$ we
introduce a fresh atom $z'$. Next, in each clause $c$ we replace some of 
its positive literals $a$ with $\neg a'$, and some of its negative 
literals $\neg b$ with $b'$ so that the resulting clause, we will denote 
it by $\hat{c}$, is the disjunction of exactly one atom and two negated 
atoms. Such replacements can always be found.

Finally, we introduce one more fresh atom, say $f$, and define $F(\vph)$
as follows:
\begin{eqnarray*}
F(\vph) & = & \{ z\vee z' \st z\in X\} \cup \{f\vee \neg z \vee\neg z' \st
z\in X\}\cup \{\neg f\} \cup \\
&& \{ \hat{c}\st c\mbox{\ is a clause in\ } \vph\}
\end{eqnarray*}
It is evident that $F(\vph)$ is in the class of theories under consideration.

We will show that $\vph$ has a model if and only if $F(\vph)$ has a model.
To this end, we note that models of $F(\vph)$ (if exist) are of the form
$M\cup \{z'\st z\in X\setminus M\}$, where $M\subseteq X$. It is now
easy to see that $M$ is a model of $\vph$ if and only if $M\cup \{z'\st
z\in X\setminus M\}$ is a model of $F(\vph)$. Thus, the claim and,
consequently, the assertion, follows. 

As we noted, the argument for the
remaining classes is similar. We only need to change the definition of 
$\hat{c}$ and use clauses $\neg z \vee\neg z'$ instead of $f\vee \neg z 
\vee\neg z'$ (there is no need to introduce $f$, as clauses being the 
disjunctions of two negated atoms are allowed in formulas in each of 
the classes considered in (1), (2) and (4)).
\hfill$\Box$

We will now use Lemma \ref{lem2} to establish the NP-hardness
of the EAS problem for $\calF(\Delta)$ for several simple sets $\Delta
\subseteq\calA$.

\begin{lemma}
\label{lem1}
If $\Delta$ is any of:
\begin{enumerate}
\item $\{[1,0,1]\}$
\item $\{[2,0,0],[0,0,1]\}$
\item $\{[3,0,0],[0,2,0]\}$
\item $\{[2,1,0],[0,2,0]\}$ 
\item $\{[2,0,0],[1,2,0],[0,1,0]\}$
\item $\{[2,0,0],[0,3,0]\}$
\end{enumerate}
then the EAS problem for $\calF(\Delta)$ is NP-hard.
\end{lemma}
Proof: (1) The proof of the NP-completeness of the EAS problem
for normal logic programs given by Marek and Truszczy\'nski \cite{mt88}
establishes the assertion (1). 

\smallskip
\noindent
(2) We will construct a reduction from the SAT problem concerning the 
class considered in Lemma \ref{lem2}(4). Let 
$\vph$ be a CNF of the 
appropriate form. We denote by $pos(\vph)$ the set of all clauses in 
$\vph$ that are of the form $a\vee b$, where $a,b\in\At$. We denote by 
$neg(\vph)$ the set of all remaining clauses in $\vph$ (all of them are 
disjunctions of at most three negative literals).

For every clause $c=\neg y_1\vee\ldots\vee \neg y_k$ in $neg(\vph)$ (where,
as we pointed out, $k\leq 3$), we introduce a fresh atom $x_c$. Next, we 
define
\begin{eqnarray*}
P(\vph) & = & \{ a | b \lla \st a\vee b \in pos(\vph)\}
\cup\\
&& \{ x_c | y_i\lla \st c\in neg(\vph),\ \mbox{$c=\neg y_1\vee\ldots\vee \neg
y_k$, $1\leq i\leq k$}\},\\
Q(\vph) & = & \{\lla \n x_c \st c\in neg(\vph)\},\ \mbox{and}\\
R(\vph) & = & P(\vph)\cup Q(\vph).
\end{eqnarray*}
To simplify notation we will write $P$, $Q$ and $R$ for $P(\vph)$,
$Q(\vph)$ and $R(\vph)$, respectively.

%
We will now show that $\vph$ is satisfiable if and only if $R$ has an 
answer set. Since $R\in \calF(\Delta)$, by Lemma \ref{lem2}(4) the assertion
will follow.

\smallskip
\noindent
($\Leftarrow$)
Let $M$ be an answer set of $R$. It follows that $M$ is an answer set 
of $P$ and so, a minimal model of $P$. Let $c$ be a clause in $\vph$.
There are two cases. First, $c=a\vee b$, where $a,b\in \At$. In that 
case, $a | b\lla\;$ is a rule in $P$. Since $M$ is a model of $P$,
$M$ is a model of $a | b\lla\;$ and, consequently, of $c$. Second, 
$c=\neg y_1\vee\ldots\vee \neg y_k$. We observe that $x_c\in M$ ($M$
being an answer set of $R$ is a model of $Q$). Since $M$ is a minimal
model of $P$, there is $j$, $1\leq j\leq k$, such that $y_j\notin M$
(otherwise $M\setminus \{x_c\}$ would be a model of $P$, contradicting
the minimality of $M$). Thus, $M$ is a model of $c$ also in this case.
Consequently, $\vph$ is satisfiable.

\smallskip
\noindent
($\Rightarrow$)
Let us assume that $\vph$ has a model, say $M$. Let $M'\subseteq M$
be a minimal model of $\vph$. It follows that $M'$ is also a minimal
model of $pos(\vph)$, as every subset of $M'$ is a model of $neg(\vph)$.
Let us define $M''=M'\cup\{x_c\st
c\in neg(\vph)\}$. Let $r$ be a clause in $P$. If $r=a | b\lla\;$, 
then $a \vee b \in pos(\vph)$ and so, $M''$ is a model of $r$. Otherwise,
$r=x_c | y_i\lla\;$, where $c=\neg y_1\vee\ldots\vee \neg y_k$ is from
$neg(\vph)$ and $1\leq i\leq k$. Since $x_c\in M''$, $M''$ is a model 
of $r$. Thus, $M''$ is a model of $P$.

Let $N\subseteq M''$ be a model of $P$. Then $N'=N\setminus\{x_c\st
c\in neg(\vph)\}$ is a model of every rule of the form $a | b\lla\;$
in $P$ and, consequently, $N'$ is a model of $pos(\vph)$. Since $N'
\subseteq M'$ and $M'$ is a minimal model of $pos(\vph)$, $N'=M'$.
Let $c\in neg(\vph)$, say $c=\neg y_1\vee\ldots\vee \neg y_k$. Since $M'$
is a model of $c$, there is $i$, $1\leq i\leq k$, such that $y_i\notin
M'$. Consequently, $y_i\notin N$. As $N$ is a model of $P$, $x_c\in N$.
Thus, $\{x_c\st c\in neg(\vph)\}\subseteq N$. It follows that $N=M''$,
that is, $M''$ is a minimal model of $P$. Since $P=\sov{R}$ and $P^{M''}=P$,
$M''$ is an answer set of $\sov{R}$. Moreover, $M''$ is a model of $Q$
(by the definition of $M''$) and $Q=\dov{R}$.
Thus, $M''$ is an answer set of $R$.

\smallskip
\noindent
(3)-(6)
In all the remaining cases, we exploit the fact that $P\in\calF(\Delta)$
has an answer set if and only if $P$ has a model (the same argument that 
we used in the proof of Lemma \ref{lem22} applies). The latter problem for
each of the cases (3)-(6) can be shown to be equivalent to the satisfiability 
problem for
the classes considered in Lemma \ref{lem2}(1)-(4), respectively. In each of
these cases the problem is NP-hard (Lemma \ref{lem2}), and so the assertion
follows. 
\hfill$\Box$

The next lemma establishes conditions guaranteeing $\Sigma_2^P$-hardness 
of the EAS problem. Eiter and Gottlob \cite{eg95} proved that given
$P\in\calF(\{[2,0,0],[1,3,0],[1,0,1]\})$, it is $\Sigma_2^P$-hard to
decide whether $P$ has an answer set. The proof can be modified to
the case when the class of input programs is restricted to
$\calF(\{[2,0,0],[1,2,0],[1,0,1]\})$, as clauses of the arity $[1,3,0]$
can be simulated by clauses of arity $[1,2,0]$. Moreover, in the 
construction provided by Eiter and Gottlob, the only rule of the arity 
$[1,0,1]$ used is of the form $x \lla \n x$, and it can be simulated by 
the rule (constraint) $\;\lla \n x$, which has the arity $[0,0,1]$. Thus, the 
$\Sigma_2^P$-hardness holds also for the class 
$\calF(\{[2,0,0],[1,2,0],[0,0,1]\})$ of programs. We omit the details
and state the result only.

\begin{lemma}
\label{lem23}
If $\Delta$ is any of:
\begin{enumerate}
\item $\{[2,0,0],[1,2,0],[0,0,1]\}$
\item $\{[2,0,0],[1,2,0],[1,0,1]\}$
\end{enumerate}
then the EAS problem for 
$\calF(\Delta)$ is $\Sigma_2^P$-hard.
\end{lemma}

We will now derive the main result of this section. It provides a complete
characterization of the complexity of the EAS problem for the class
$\calF(\Delta)$. To state the result we 
introduce one more piece of notation. Given $\Delta,
\Theta \subseteq \calA$, we write $\Delta\preceq\Theta$ if
for every $\alpha\in\Delta$ there is $\beta\in\Theta$ such that 
$\alpha\preceq\beta$. For instance, we have $\{[1,\infty,1], \{[\infty, 0,0]\}
\preceq \{[\infty,\infty, 1]\}$. Clearly, if $\Delta \preceq \Theta$ then 
$\calF(\Delta) \subseteq \calF(\Theta)$. We will use this property
frequently in proofs throughout the paper.

\begin{theorem}
\label{thm:main}
Let $\Delta\subseteq\calA$ be finite.

\smallskip
\noindent
(A) If 
\begin{enumerate}
\item $\Delta\preceq\{[\infty,\infty,0],[0,0,0]\}$, or 
\item $\Delta\preceq\{[1,\infty,0],[0,\infty,\infty]\}$, or
\item $\Delta\preceq\{[\infty,1,0],[0,1,0]\}$, or 
\item $\Delta\preceq\{[i,j,0] \in\calA\st i+j\leq 2\}$,
\end{enumerate}
then the EAS problem for $\calF(\Delta)$ is in $\mathrm{P}$.

\smallskip
\noindent
(B) Otherwise, if 
\begin{enumerate}
\item $\Delta\preceq\{[1,\infty,\infty],[0,\infty,\infty]\}$, or
\item $\Delta\preceq\{[\infty,1,\infty],[0,\infty,\infty]\}$, or 
\item $\Delta\preceq\{[\infty,\infty,0],[0,\infty,0]\}$,
\end{enumerate}
then the EAS problem for $\calF(\Delta)$ is $\mathrm{NP}$-complete.

\smallskip
\noindent
(C) Otherwise, the EAS problem for $\calF(\Delta)$ is $\Sigma_2^P$-complete.
\end{theorem}
Proof: The claim (A) follows directly from Lemma \ref{lem21}. Thus, 
let us assume 
that $\Delta$ does not fall under the scope of (A) and satisfies the
assumptions of (B). By Lemma \ref{lem22}, the latter implies that the
EAS problem for $\calF(\Delta)$ is in NP. 

If $\{[1,0,1]\}\preceq\Delta$ or $\{[2,0,0],[0,0,1]\}\preceq\Delta$,
the NP-hardness of the EAS problem for $\calF(\Delta)$ follows from
Lemma \ref{lem1}, parts (1) and (2), respectively. Thus, let us assume that 
$\{[1,0,1]\}\not\preceq
\Delta$ and $\{[2,0,0],[0,0,1]\}\not\preceq\Delta$.

Since $\{[1,0,1]\}\not\preceq\Delta$, we have $\Delta\preceq\{[\infty,
\infty,0],$ $[0,\infty,\infty]\}$. Since $\Delta$ does not satisfy the 
condition (A2), $\{[2,0,0]\}\preceq\Delta$. Since $\{[2,0,0],[0,0,1]\}
\not\preceq\Delta$, $\{[0,0,1]\}\not\preceq\Delta$. Thus, $\Delta\preceq
\{[\infty,\infty,0],[0,\infty,0]\}$. 

Since $\Delta$ does not satisfy the condition (A1), $\{[0,1,0]\}\preceq
\Delta$. Similarly, since $\Delta$ does not satisfy the condition (A3), 
$\{[1,2,0]\}\preceq \Delta$ or $\{[0,2,0]\}\preceq \Delta$. We also have
that $\Delta$ does not satisfy the condition (A4). Thus, there is 
$\alpha\in\Delta$ such that $\alpha_1+\alpha_2\geq 3$. Since we already
proved that $\{[2,0,0]\}\preceq\Delta$, it follows that at least one of 
the following conditions holds:
$\{[3,0,0],[0,2,0]\}\preceq\Delta$,
$\{[2,1,0],[0,2,0]\}\preceq\Delta$,
$\{[2,0,0],[1,2,0],[0,1,0]\}\preceq\Delta$, or
$\{[2,0,0],[0,3,0]\}\preceq\Delta$.
Thus, the NP-hardness of the EAS problem for $\calF(\Delta)$ follows 
again from Lemma \ref{lem1} and completes the proof of (B).

To prove (C), let us assume that $\Delta$ does not fall under 
the scope of (B). Since $\Delta$ does not satisfy (B1), $\{[2,0,0]\}
\preceq\Delta$. Similarly, since $\Delta$ does not satisfy (B2),
$\{[1,2,0]\}\preceq\Delta$. Finally, since $\Delta$ does not 
satisfy (B3), $\{[0,0,1]\}\preceq\Delta$ or $\{[1,0,1]\}\preceq\Delta$.
Thus, the $\Sigma_2^P$-hardness follows by Lemma \ref{lem23}. Since the 
EAS problem is in $\Sigma_2^P$ even without any restrictions on the class
of programs, both (C) and the assertion of the lemma follows.
\hfill$\Box$

\section{The Case of Infinite $\Delta$}
\label{infinite}

The question we study now is whether there are interesting classes of
programs of the form $\calF(\Delta)$, when $\Delta$ is infinite. The 
main result of this section is that by allowing $\Delta$ to be infinite, 
we do not obtain \emph{any} new classes of programs. In other words,
for every class of programs of the form $\calF(\Delta)$ there is a 
finite set $\Delta'\subseteq\calA$ such that $\calF(\Delta)=
\calF(\Delta')$.

A sequence $\{\alpha^k\}_{k=1}^\infty$ is \emph{monotone} (\emph{strictly}
monotone) if for every $k$, $\alpha^k\preceq \alpha^{k+1}$ ($\alpha^k
\prec\alpha^{k+1}$, respectively).
Let $\{\alpha^k\}_{k=1}^\infty$ be a \emph{monotone} sequence of
elements of $\calA$.  We define the \emph{limit} of this sequence as 
$\alpha^\infty=[(\alpha^\infty)_1,(\alpha^\infty)_2,
(\alpha^\infty)_3]$, where $(\alpha^\infty)_i= \sup\{(\alpha^k)_i\st 
k=1,2\ldots,\}$, for $i=1,2,3$. We stress that we do not consider transfinite
sequences here. All sequences have the set of natural numbers as their domain. 
Let $\Delta\subseteq\calA$. A monotone sequence $\{\alpha^k\}_{k=1
}^\infty$ of elements of $\Delta$ is \emph{maximal} if there is
\emph{no} $\alpha\in\Delta$ such that $\alpha^\infty\prec\alpha$. We
define $A(\Delta)$ to be the set of the limits of
maximal sequences in $\Delta$.

We have the following two lemmas (we omit the proof of the first one 
as it is evident).

\begin{lemma} 
\label{lem24}
Let $\{\beta^k\}_{k=1}^\infty$ be a strictly monotone sequence of elements 
from $\calA$. For every $\alpha\in\calA$, if $\alpha\prec\beta^\infty$,
then there is $k$ such that $\alpha\prec\beta^k$.
\end{lemma}



\begin{lemma}
\label{prop11}
Let $\Delta\subseteq\calA$ and $\alpha\in\Delta$. Then, there is
$\alpha'\in A(\Delta)$ such that $\alpha\preceq\alpha'$.
\end{lemma}
Proof: Let $X=\{\beta\in\Delta\st \alpha\preceq\beta\}$. If $X$
has a maximal element, say $\gamma$, a sequence with each term equal 
to $\gamma$ is maximal. Its limit, also equal to $\gamma$, clearly 
satisfies $\gamma\in A(\Delta)$ and $\alpha\preceq\gamma$. Thus, the 
assertion follows.

Otherwise, $X$ has no maximal elements. Let $\alpha^1$ be any element 
in $X$ (we note that $X\not=\emptyset$, as $\alpha\in X$). Let $k\geq 1$
and let $\langle \alpha^1,\ldots,\alpha^k\rangle$ be a strictly monotone
sequence of
$k$ elements in $X$, for some $k\geq 1$. Since $X$ has no maximal 
elements, $X$ contains elements that are strictly greater than $\alpha^k$.
Let us select as $\alpha^{k+1}$ an element $\beta\in X$ such that
$\alpha^k\prec \beta$ and $\alpha^k_i < \beta_i$ holds on as many positions
$i=1,2,3$ as possible. An infinite sequence we define in this way, we will 
denote it by $\alpha$, is strictly monotone. Let us assume that there
is $\beta\in \Delta$ such that $\alpha^\infty \prec \beta$. It follows
that there is $j$, $1\leq j \leq 3$, such that $(\alpha^\infty)_j<\beta_j$.
Thus, $(\alpha^\infty)_j =m$, for some integer $m$, and there is $n$ such
that $(\alpha^n)_j=m$.
Since $\alpha^{n+1} \prec \beta$ and $(\alpha^{n})_j=(\alpha^{n+1})_j=m$,
the number of positions $i$ such that $(\alpha^n)_i < (\alpha^{n+1})_i$
is strictly smaller than the number of positions $i$ such that 
$(\alpha^n)_i < \beta_i$. Since $\beta \in X$, that contradicts the way
we constructed the sequence $\alpha$.

It follows that the sequence $\{\alpha^k\}_{k=1}^\infty$ is maximal for 
$\Delta$ and so, the assertion follows in this case, too. \hfill$\Box$

We now have the following properties. 

\begin{proposition}
\label{prop7b}
For every $\Delta\subseteq\calA$, $\calF(\Delta)=\calF(A(\Delta))$.
\end{proposition}
Proof: To prove the assertion, it is enough to show that for every
\emph{arity} $\alpha$ (no occurrence of $\infty$), $\{\alpha\}\preceq
\Delta$ if and only if $\{\alpha\}\preceq A(\Delta)$. Let us first
assume that $\{\alpha\}\preceq \Delta$. It follows that there is an
element $\alpha'\in\Delta$ such that $\alpha\preceq\alpha'$. By
Lemma \ref{prop11}, there is $\alpha''\in A(\Delta)$ such that
$\alpha'\preceq\alpha''$. Thus, $\alpha\preceq\alpha''$ and so,
$\{\alpha\}\preceq A(\Delta)$.

Conversely, let  $\{\alpha\}\preceq A(\Delta)$. It follows that there
is $\beta\in A(\Delta)$ such that $\alpha\preceq\beta$. Since $\beta\in
A(\Delta)$, there is a monotone sequence $\{\beta^k\}_{k=1}^\infty$ of
elements of $\Delta$ such that its limit is $\beta$. Without loss of
generality we can assume 
that either
starting with some $k_0$, the sequence $\{\beta^k\}_{k=1}^\infty$
is constant, or
the sequence $\beta$ is strictly monotone.
In the first case, $\alpha\preceq\beta=\beta^{k_0}$. Since $\beta^{k_0}
\in\Delta$, $\{\alpha\}\preceq \Delta$. In the second case, Lemma 
\ref{lem24} implies that there is $k$ such that $\alpha\prec\beta^k$,
and again $\{\alpha\}\preceq \Delta$ follows.
\hfill$\Box$

\begin{proposition}
\label{prop7a}
For every $\Delta$, $A(\Delta)$ is an antichain.
\end{proposition}
Proof: Let us assume that $A(\Delta)$ is not an antichain. Then there
are $\alpha,\alpha'\in A(\Delta)$ such that $\alpha\prec\alpha'$. We have
that $\alpha'\notin\Delta$ (otherwise, the sequence that $\alpha$ is a 
limit of would not be maximal). Consequently, $\alpha'$ is the limit
of a strictly monotone sequence of elements from $\Delta$. By Lemma \ref{lem24},
there is an element $\beta$ in the sequence such that $\alpha\prec\beta$.
That contradicts the fact that $\alpha$ is the limit of a maximal sequence
and yields the assertion. \hfill$\Box$

\begin{proposition} \label{prop3}
Every antichain in the partially ordered set $\langle \calA,\preceq
\rangle$ is finite.
\end{proposition}
Proof: We will first prove that every sequence $s=\{[b^k,c^k]\}_{k=1}^\infty$
such that (i) $b^k,c^k\in U$, for $k=1,2,\ldots$, and (ii) the set of 
distinct elements occurring in the sequence is infinite, contains an 
infinite strictly monotone subsequence. 

We start by observing that by (ii) one can select an infinite
subsequence of $s$, 
in which all elements are distinct. Thus, without loss of generality, we 
may assume that, in fact, all elements in $s$ are distinct. If there is 
$b$ such that $C(b)=\{k\st b^k =b\}$ is infinite, then the assertion is 
evident. Indeed, the sequence $\{c^k\}_{k\in C(b)}$ contains no repetitions
(as $s$ contains no repetitions) and, consequently, contains a strictly
increasing subsequence. Thus, let us assume that for every $b$, the
set $C(b)$ is finite. It follows that the sequence $\{b^k\}_{k=1}^\infty$
contains a strictly increasing subsequence, say $\{b^{k_j}\}_{j=1}^\infty$. 

If there is $c$ that occurs in the corresponding sequence $\{c^{k_j}\}_{j=
1}^\infty$ infinitely many times, then these occurrences yield a strictly 
monotone subsequence of $\{[b^{i_j},c^{i_j}]\}_{j=1}^\infty$ and, consequently,
a strictly monotone subsequence of $\{[b^k,c^k]\}_{k=1}^\infty$. Otherwise,
$\{c^{k_j}\}_{j=1}^\infty$ contains infinitely many elements. Thus, it contains a strictly monotone subsequence which, together with the corresponding $b$'s,
yields a strictly monotone subsequence of $\{[b^k,c^k]\}_{k=1}^\infty$. 

To prove the assertion of the proposition, let us assume 
that there is an \emph{infinite} antichain $A$ in $\langle \calA,\preceq
\rangle$. Let $\{\alpha^k\}_{k=1}^\infty$ be any enumeration of the
elements of $A$ (without repetitions). If there is $a \in U$ such that 
$A_a=\{\alpha\in A\st \alpha_1=a\}$ is infinite, then the fact proved
above yields a contradiction. 

Thus, for every 
$a\in U$, the set $A_a$ is finite. Since $A$ is infinite, there is an 
infinite subsequence $\beta$ of $\alpha$ such that for every $n=1,2,\ldots$,
$(\beta^n)_1 < (\beta^{n+1})_1$. 
As $A$ is an antichain, the sequence $\{[(\beta^n)_2,(\beta^n)_3]\}_{n=
1}^\infty$ contains no repeating elements. By the property proved above,
it contains an infinite strictly monotone subsequence, a contradiction.  
\hfill$\Box$

These properties imply the main result of this section. It asserts
that every class of programs $\calF(\Delta)$ can be defined by means of 
a finite set $\Delta'$ that is an antichain in $\langle \calA,
\preceq\rangle$.

\begin{theorem}
\label{cor11}
For every set $\Delta\subseteq\calA$ there is a finite subset
$\Delta'\subseteq\calA$ such that $\Delta'$ is an antichain and 
$\calF(\Delta)= \calF(\Delta')$.
\end{theorem}
Proof: Let us define $\Delta'=A(\Delta)$. By Propositions \ref{prop7a}
and \ref{prop3}, $\Delta'$
is a finite antichain in $\langle \calA, \preceq\rangle$, and by
Proposition \ref{prop7b}, $\calF(\Delta)= \calF(\Delta')$. Thus, 
the theorem follows.
\hfill$\Box$ 

\section{The Complexity of Skeptical and Credulous Reasoning}

The EAS problem is just one example of a reasoning task that arises in the 
context of disjunctive logic programs with the answer-set semantics. 
There are several other tasks that are of interest, too. They concern
deciding whether a program nonmonotonically entails a literal, that is
an atom, say $a$, or its negation $\neg a$. 

We recall that if $M$ is a set of atoms (an interpretation) and $a$ is 
an atom, then $M\models a$ if $a \in M$, and $M\models\neg a$ if $a\notin
M$. For a disjunctive logic program $P$ and a literal $l$ we say that
\begin{enumerate}
\item $P$ \emph{skeptically entails} $l$, written $P\models_{s} l$, if 
$M\models l$, for every answer set $M$ of $P$;
\item $P$ \emph{credulously entails} $l$, written $P\models_{c} l$, if 
there is an answer set $M$ of $P$ such that $M\models l$.
\end{enumerate}

We note that $P\models_s l$ if and only if $P\not\models_c \overline{l}$,
where $\overline{l}$ is $l$'s \emph{dual} literal. Thus, to establish fully
the complexity of deciding nonmonotonic entailment it is enough to focus
on deciding whether $P\models_c \neg a$ and $P\models_s \neg a$, where
$a$ is an atom. These two decision tasks were studied by Eiter and Gottlob 
\cite{eg95}, who proved that, in general, the first one is 
$\Sigma_2^P$-complete and the second one is $\Pi_2^P$-complete. 

Reasoning with answer sets is related to circumscription and closed-world 
reasoning with propositional theories. A detailed study of the complexity
of those forms of reasoning was conducted by Cadoli and Lenzerini 
\citeyear{cl94}. Using Theorem \ref{thm:main} and one of the results from 
that paper (which we state in the proof below), one can characterize in terms 
of our definition schema the complexity of deciding, given 
a program $P$ and an atom $a$, whether $P\models_c \neg a$ and $P\models_s
\neg a$. The two problems are addressed in the following two theorems.

\begin{theorem}
\label{thm:cr}
Let $\Delta\subseteq\calA$ be finite.

\smallskip
\noindent
(A) If
\begin{enumerate}
\item $\Delta\preceq\{[1,\infty,0],[0,\infty,\infty]\}$, or
\item $\Delta\preceq\{[\infty,1,0],[0,1,0]\}$, or
\item $\Delta\preceq\{[i,j,0] \in\calA\st i+j\leq 2\}$,
\end{enumerate}
then the problem to decide whether $P\models_c \neg a$, where $P\in 
\calF(\Delta)$ and $a$ is an atom, is in $\mathrm{P}$.

\smallskip
\noindent
(B) Otherwise, if
\begin{enumerate}
\item $\Delta\preceq\{[1,\infty,\infty],[0,\infty,\infty]\}$, or
\item $\Delta\preceq\{[\infty,1,\infty],[0,\infty,\infty]\}$, or
\item $\Delta\preceq\{[\infty,\infty,0],[0,\infty,0]\}$,
\end{enumerate}
then the problem to decide whether $P\models_c \neg a$, where $P\in 
\calF(\Delta)$ and $a$ is an atom, is $\mathrm{NP}$-complete.

\smallskip
\noindent
(C) Otherwise, the problem to decide whether $P\models_c \neg a$, where 
$P\in\calF(\Delta)$ and $a$ is an atom, is $\Sigma_P^2$-complete.
\end{theorem}
Proof: It is well known that $P$ has an answer set $M$ such that 
$M\models \neg a$ (that is, $a\notin M$) if and only if $P\cup
\{\;\leftarrow a\}$ has an answer set. Let $\Delta\subseteq\calA$ be
finite and let us define $\Delta'=\Delta\cup\{[0,1,0]\}$. Clearly, if
$P\in\calF(\Delta)$, then $P\cup\{\;\leftarrow a\}\in \calF(\Delta')$.
Moreover, if $\Delta$ falls under the scope of (A) ((A) or (B), respectively)
of this theorem then $\Delta'$ falls under the scope of (A) ((A) or (B), 
respectively) of Theorem \ref{thm:main}. Consequently, the upper bound
follows by Theorem \ref{thm:main}.

To prove the lower bounds, we first consider the case when $\Delta$ 
does not satisfy any of the conditions listed in (A) and $\Delta\preceq
\{[\infty,\infty,0],[0,0,0]\}$ (thus, in particular, $\Delta$ falls under
the scope of (B)). Since $\Delta$ does not fall under the condition (A1),
$\{[2,0,0]\}\preceq\Delta$. Similarly, since $\Delta$ does not fall under 
the condition (A2), $\{[1,2,0]\}\preceq\Delta$.

Let $P\in\calF(\Delta')$, where $\Delta'=
\{[2,0,0],[1,2,0],[0,1,0]\}$. We select a fresh atom $a$ and define
$P'=\sov{P}\cup\{a \leftarrow \bd(r)\st r\in\dov{P}\}$. Clearly, $P'\in
\calF(\Delta)$. Moreover, $P$ has an answer set if and only if $P'\models_c 
\neg a$. Thus, the claim follows by Lemma \ref{lem1}(5).

From now on, we assume that $\Delta\not\preceq\{[\infty,\infty,0],[0,0,0]\}$ 
and $\Delta$ does not satisfy any of the conditions listed in (A). Let 
us also assume that $\Delta$ falls under the scope of (B) or (C) of this 
theorem. It follows that $\Delta$ falls under the scope of the corresponding
case of Theorem \ref{thm:main}. 

If $\Delta$ satisfies (B1), then $\{[1,0,1]\}\preceq\Delta$ (otherwise,
(A1) would hold). Let $P\in\calF(\Delta)$. Let $a,b$ be fresh atoms and 
let us set $P'=P\cup\{a\leftarrow \n b; b\leftarrow \n a\}$. 
If $\Delta$ does not satisfy (B1), $\{[2,0,0]\} \preceq\Delta$.                 
Let $P\in\calF(\Delta)$, and let us set $P'=P\cup\{a | b\}$, where $a,b$ are
fresh atoms. Clearly, in either case, $P'\in\calF(\Delta)$, and $P$ has an 
answer set if and only if $P'\models_c \neg a$. Thus, all the 
remaining lower bounds follow from Theorem \ref{thm:main}.
\hfill$\Box$.

\begin{theorem}
\label{thm:sk}
Let $\Delta\subseteq\calA$ be finite.

\smallskip
\noindent
(A) If $\Delta\preceq\{[1,\infty,0],[0,\infty,\infty]\}$, 
then the problem to decide whether $P\models_s \neg a$, where $P\in
\calF(\Delta)$ and $a$ is an atom, is in $\mathrm{P}$.

\smallskip
\noindent
(B) Otherwise, if
\begin{enumerate}
\item $\Delta\preceq\{[1,\infty,\infty],[0,\infty,\infty]\}$, or
\item $\Delta\preceq\{[\infty,1,\infty],[0,\infty,\infty]\}$,
\end{enumerate}
then the problem to decide whether $P\models_s \neg a$, where $P\in
\calF(\Delta)$ and $a$ is an atom, is $\mathrm{coNP}$-complete.

\smallskip
\noindent
(C) Otherwise, the problem to decide whether $P\models_s \neg a$, where
$P\in\calF(\Delta)$ and $a$ is an atom, is $\Pi^P_2$-complete.
\end{theorem}
Proof: It is well known that $P$ has an answer set such that
$M\not\models\neg a$ (that is, $a\in M)$ if and only if $P\cup
\{\;\leftarrow \n a\}$ has an answer set. That 
observation implies all upper bound results (by a similar argument as that
used in the proof of the previous theorem).

We will now prove the lower bounds for the cases (B) and (C). Let us
assume that $\Delta$ does not satisfy (A) but falls under the scope
of (B). If $\Delta$ satisfies (B1),
then $\{[1,0,1]\}\preceq\Delta$. Let $P\in\calF(\Delta)$ and let $a$ and
$a'$ be fresh atoms. We note that $P$ has an answer set if and only if $P\cup
\{a\leftarrow \n a'\}$ has an answer set $M$ such that $a\in M$ or,
equivalently, if and only if $P\cup \{a\leftarrow \n a'\} \not\models_s
\neg a$. Thus, the hardness follows (cf. Lemma \ref{lem1}(1)).

Let us assume then that $\Delta$ does not satisfy (B1). Then, we have that
$\{[2,0,0]\}\preceq\Delta$. It follows from the results of Cadoli and
Lenzerini \citeyear{cl94} that it is NP-complete to decide whether a given
2CNF theory whose every clause is a disjunction of two atoms has a minimal
model that contains a given atom $a$.
As minimal models of such theories are precisely answer sets of the 
corresponding disjunctive program, it follows that given a program 
$P\in\calF(\Delta)$ and an atom $a$, it is coNP-hard to decide whether 
$P\models_s \neg a$.

Finally, let us assume that neither (A) nor (B) apply to $\Delta$. Then
$[1,2,0]\in\Delta$ and $[2,0,0]\in\Delta$. Eiter and Gottlob \cite{eg95}
proved that if $P\in\calF(\{[2,0,0],[1,3,0]\})$ and $a$ is an atom then
it is $\Pi_2^P$-hard to decide whether $P\models_s \neg a$. That result
can be strengthened to the case when $P\in\calF(\{[2,0,0],[1,2,0]\})$,
as clauses of the arity $[1,3,0]$ can be simulated by clauses of arity 
$[1,2,0]$ (cf. the comments preceding Lemma \ref{lem23}).
\hfill$\Box$

We note that credulous reasoning is simple (in P or in NP) for several
classes of programs. In contrast, there are fewer classes of programs,
for which skeptical reasoning is simple (in P or coNP). The main reason
behind this asymmetry is that in the cases (A2), (A3) and (B3) of Theorem
\ref{thm:cr} (positive programs) answer sets and minimal models coincide. 
Thus, in these 
cases, credulous reasoning asks for the existence of a \emph{minimal} 
model that does not contain an atom $a$, which is equivalent to the 
existence of a model (not necessarily minimal) that does not contain 
$a$. In other words, the requirement of minimality becomes immaterial
(one source of complexity disappears). This is not so with skeptical 
reasoning, where not having $a$ in any minimal model is not the same as 
not having $a$ in any model. A similar comparison of skeptical 
and credulous reasoning for positive programs was offered by Eiter and
Gottlob for the coarser setting of classes of programs they considered
\cite{eg95}.

\section{Another Representation Schema --- Explicit Arities}

Next, we consider briefly an alternative way, in which classes of
programs could
be described by means of arities of rules. When defining the class $\calF
(\Delta)$, we view each element $\alpha\in\Delta$ as a shorthand for the
set of all arities $\beta$ such that $\beta\preceq\alpha$. In other 
words, $\Delta$ is an \emph{implicit} representation of the set of all 
allowed arities: not only those arities that are explicitly listed in 
$\Delta$ are legal but also those that are ``dominated'' by them.

There is another, more direct (more explicit), way
to use arities to define classes of programs. Let $\Delta\subseteq\calA$
be a set of \emph{arities}, that is, we now do not allow super\-arities in 
$\Delta$. We define $\calG(\Delta)$ to consist of all finite programs $P$
such that for every rule $r\in P$, the arity $\alpha$ of $r$ belongs to 
$\Delta$. Thus, when defining 
the class $\calG(\Delta)$, $\Delta$ serves as an \emph{explicit}
specification of the set of allowed arities.  

One can show that the results of Section \ref{finite} can be adapted to
the setting of classes of the form $\calG(\Delta)$, where $\Delta$ is a
set of arities. In particular, we have the following 
result.

\begin{theorem}
\label{thm:ex}
Let $\Delta\subseteq\calA$ be a 
set of arities. If there are no
$k\geq 1$ and $m\geq 0$ such that $\{[k,0,m]\}\in\Delta$,
then the EAS problem for $\calG(\Delta)$ is in $\mathrm{P}$. Otherwise:

\smallskip
\noindent
(A) If
\begin{enumerate}
\item $\Delta\preceq\{[\infty,\infty,0],[0,0,0]\}$, or
\item $\Delta\preceq\{[1,\infty,0],[0,\infty,\infty]\}$, or
\item $\Delta\preceq\{[\infty,1,0],[0,1,0]\}$, or
\item $\Delta\preceq\{[i,j,0] \in\calA\st i+j\leq 2\}$,
\end{enumerate}
then the EAS problem for $\calG(\Delta)$ is in $\mathrm{P}$.

\smallskip
\noindent
(B) Otherwise, if
\begin{enumerate}
\item $\Delta\preceq\{[1,\infty,\infty],[0,\infty,\infty]\}$, or
\item $\Delta\preceq\{[\infty,1,\infty],[0,\infty,\infty]\}$, or
\item $\Delta\preceq\{[\infty,\infty,0],[0,\infty,0]\}$,
\end{enumerate}
then the EAS problem for $\calG(\Delta)$ is $\mathrm{NP}$-complete.

\smallskip
\noindent
(C) Otherwise, the EAS problem for $\calG(\Delta)$ is 
$\Sigma_P^2$-complete.
\end{theorem}
Proof (sketch): Let us first assume that there are no $k\geq 1$ and $m\geq 0$
such that $\{[k,0,m]\}\in\Delta$, and let $P\in\calG(\Delta)$. Then every 
rule in $\sov{P}$ has at least one positive 
atom in the body and so, $M=\emptyset$ is the unique answer set of
$\sov{P}$. It can be verified in polynomial time whether $M=\emptyset$
is a model of $\dov{P}$. Thus, the EAS problem for programs in 
$\calG(\Delta)$ can be decided in polynomial time.

To prove the remaining part of the assertion, we note that the upper
bound is implied directly by Theorem \ref{thm:main} (as $\calG(\Delta)
\subseteq\calF(\Delta)$). To prove the lower bounds, we observe that 
if there are $k\geq 1$ and $m\geq 0$ such that $\{[k,0,m]\}\in\Delta$,
then the EAS problem for $\calF(\Delta)$ can be reduced to
the EAS problem for $\calG(\Delta)$. Indeed, let $P\in \calF(\Delta)$
and let $r\in P$. Then there is $\alpha\in\Delta$ such that $\alpha_r\preceq
\alpha$, where $\alpha_r$ is the arity of $r$. Having 
$\{[k,0,0]\}\in\Delta$, where $k\geq 1$, or $\{[k,0,m] \}
\in\Delta$, where
$k,m\geq 1$, allows us to ``simulate'' the effect of $r$ with a rule $r'$
of arity $\alpha$ obtained by repeating atoms in the head of $r$, and by
inserting an atom $a$ and a negated atom $\n b$, where $a$ and $b$ are
fresh, as many times as necessary in the body of $r$ to ``reach'' the 
arity $\alpha$. We also add the rule $a|\ldots|a\leftarrow$ or $a|
\ldots|a\leftarrow \n a',\ldots,\n a'$, where $a'$ is another fresh
atom and $a$ and $\n a'$ are repeated $k$, or $k$ and $m$ times, 
respectively.
\hfill$\Box$

%

\section{The Case of Supported Models}

Lastly, we will now consider the problem of the existence of 
supported models,
the \emph{ESPM} problem, for short. Let $M$ be a set of atoms. A logic 
program rule $r$ is \emph{$M$-applicable} if $M$ satisfies the body of
$r$ (that is, satisfies every literal in the body). For a logic program 
$P$ and a set of atoms $M$, we define $H(P,M)$ to consist of the heads
of all rules in $\sov{P}$ that are $M$-applicable. We say that a set of
atoms $M$ is a \emph{supported} model of $P$ if $M$ is a minimal model 
of $H(P,M)$ and a model of $\dov{P}$ \cite{bradix96jlp1,Inoue98-JLP}.
The following properties of supported models are well known.

\begin{enumerate}
\item[S1.] Every supported model of a program $P$ is a model of $P$.
\item[S2.] If $M$ is a supported model of $P$ then every atom in $M$ 
occurs in the head of some $M$-applicable rule in $P$ and is the 
\emph{only} atom from $M$ in the head of that rule.
\item[S3.] Every answer set of a program $P$ is a supported model of $P$.
\item[S4.] If all rules in $\sov{P}$ are \emph{purely negative} (their bodies
contain
no non-negated occurrences of atoms), then every supported model of $P$ 
is an answer set of $P$.
\item[S5.] If $P$ is a finite proper Horn program, then $P$ has the greatest 
supported model, and it can be computed in polynomial time.
\item[S6.] The ESPM problem is NP-complete.
\end{enumerate}

The main result of this section is a \emph{dichotomy} result for the ESPM
problem (by the property (S6), even for the general class of all disjunctive
programs, the ESPM problem is in the class NP). Specifically, we will show
that for every set $\Delta\subseteq\calA$ the ESPM
problem for programs in $\calF(\Delta)$ is either in P or is NP-complete,
and we will characterize sets $\Delta$, for which the former holds. The
proof relies on a series of lemmas providing upper and lower bounds for
the complexity of the ESPM problem for programs in $\calF(\Delta)$ for 
some particular sets $\Delta\subseteq\calA$. We note that thanks to our
results from Section \ref{infinite}, it is enough to restrict attention 
to the case of finite sets $\Delta$ only. Thus, we adopt this assumption 
here.

We start with a lemma that, for some programs, connects the problem of 
the existence of a supported model to that of the existence of a model
(and so, to the SAT problem).

\begin{lemma}
\label{lem:supp1}
Let $P\in\calF(\{[\infty,\infty,0],[0,\infty,0]\})$. Then $P$ has a 
supported model if and only if $P$ has a model.
\end{lemma}
Proof: Thanks to the property (S1) above, only the ``if'' part needs 
to be proved. Thus, let us assume that $M$ is a model of $P$. It follows that 
$M$ is a model of $\sov{P}$. Let $M'$ be a minimal model of $\sov{P}$
such that $M'\subseteq M$ (its existence follows by the finiteness of
$P$, cf. the definition of $\calF(\Delta)$). Then $M'$ is an answer set
 of $\sov{P}$ and, by the property (S3), a supported model of $\sov{P}$.
Moreover, since $\dov{P}\in\calF(\{[0,\infty,0]\})$ and $M$ is a model 
of $\dov{P}$, $M'$ is a model of $\dov{P}$. Thus, $M'$ is a supported 
model of $P$.
\hfill$\Box$

Next, we present and prove a lemma that exhibits classes of programs, 
for which the ESPM problem is in the class P.

\begin{lemma}
\label{lem:supp:upper}
Let $\Delta$ satisfy at least one of the following conditions:
\begin{enumerate}
\item $\Delta\preceq\{[1,0,0],[0,\infty,\infty]\}$
\item $\Delta\preceq\{[\infty,\infty,0],[0,0,0]\}$
\item $\Delta\preceq\{[\infty,1,0],[0,1,0]\}$
\item $\Delta\preceq\{[1,\infty,0],[0,\infty,0]\}$
\item $\Delta\preceq\{[2,0,0],[1,1,0],[0,2,0]\}$
\item $\Delta\preceq\{[1,\infty,0],[0,0,\infty]\}$
\item $\Delta\preceq\{[1,1,0],[0,1,\infty]\}$.
\end{enumerate}
Then the ESPM problem for programs in $\calF(\Delta)$ is in \rm{P}.
\end{lemma}
Proof: (1) If $P\in\calF([1,0,0],[0,\infty,\infty])$, then $\sov{P}$
is just a set of facts. This set of facts, say $M$, is the only supported 
model of $\sov{P}$. Thus, if $M$ satisfies the constraints in $\dov{P}$,
$M$ is a supported model of $P$. Otherwise, $P$ has no supported models.
Since checking whether a set of atoms satisfies constraints in $\dov{P}$
is a polynomial-time task, checking whether $P\in\calF([1,0,0],[0,\infty,
\infty])$ has a supported model is a polynomial-time task, too.

\smallskip
\noindent 
(2)-(5) In each of these cases, there is a polynomial-time algorithm
for testing whether $P\in\calF(\Delta)$ has a model.
Thus, the claim follows from Lemma \ref{lem:supp1}.
 
\smallskip
\noindent
(6) To decide whether $P\in\calF(\{[1,\infty,0],[0,0,\infty]\})$ has a
supported model, we proceed as follows. First, we compute the greatest
supported model of $\sov{P}$ (cf. the property (S5)). Let us denote
this supported model by $M$. Since $\dov{P}\in\calF(\{[0,0,\infty]\})$,
$\sov{P}$ has a supported model that satisfies $\dov{P}$ if and only if
$M$ satisfies $\dov{P}$. Thus, once we compute $M$, we check whether
$M$ satisfies $\dov{P}$. If so, we decide the ESPM problem for $P$ in 
positive; otherwise, we decide the ESPM problem for $P$ in negative. 
Since $M$ can be computed in polynomial time (again, the property (S5)),
the claim follows.

\smallskip
\noindent
(7) The key to a proof is a certain transformation of programs in 
$\calF(\{[1,1,0],[0,1,\infty]\})$ that does not affect the status
of the ESPM problem and runs in polynomial time. Let $P\in
\calF(\{[1,1,0],[0,1,\infty]\})$.
\begin{enumerate}
\item If ${P}$ contains a fact, say $a$, remove all proper
rules with $a$ in the head, and all constraints with a negated
occurrence of $a$; then remove $a$ from all remaining rules.
\item If there is an atom, say $a$, that occurs in $P$ but never in
the head of a rule, remove from $P$ all rules with a non-negated
occurrence of $a$ in the body and remove all negated occurrences of
$a$ from the remaining rules.
\item If there is an atom, say $a$, such that $\leftarrow a$ is a 
constraint and $a$ appears in the head of a proper rule in $P$, say
$a\leftarrow b$, remove the rule from $P$ and add the constraint 
$\leftarrow b$.
\end{enumerate} 
For each rule above, one can check that $P$ has a supported model if 
and only if
the program resulting from applying  the rule has a supported model. It
is also clear that the process of applying the steps (1)-(3) will terminate
after polynomially many iterations. Indeed, in each case, when we apply a
rule, we decrease the size of the program. It follows that the process
runs in polynomial time. Let us denote by $P'$ the program that 
results when no step is applicable.

Clearly, every atom that appears in the body of a rule in $\sov{P'}$ 
appears also in the head of a rule in $\sov{P'}$. Thus, the set of atoms, 
say $H$, that appear in the heads of rules in $\sov{P'}$ is a supported 
model of $\sov{P'}$. Moreover, for every constraint $\leftarrow a$, 
$a\notin H$, and for every atom $a$ occurring negated in a constraint, 
$a\in H$. If $P'$ contains a contradictory constraint (empty head and 
empty body) then it has no models and so no supported models. Thus, 
$P$ has no supported models either. Otherwise, $H$ satisfies the 
constraints of $P'$ and so is a supported model of $P'$. Thus, $P$ has a
supported model, too.  \hfill$\Box$

The next lemma exhibits several simple classes of programs $\calF(\Delta)$
for which the ESPM problem is NP-complete.

\begin{lemma}
\label{lem:supp:lower}
Let $\Delta$ be one of the following sets:
\begin{enumerate}
\item $\Delta=\{[1,0,1]\}$
\item $\Delta=\{[2,0,0],[0,0,1]\}$
\item $\Delta=\{[3,0,0],[0,2,0]\}$
\item $\Delta=\{[2,1,0],[0,2,0]\}$
\item $\Delta=\{[2,0,0],[1,2,0],[0,1,0]\}$
\item $\Delta=\{[2,0,0],[0,3,0]\}$
\item $\Delta=\{[1,1,0],[0,2,0],[0,0,1]\}$
\item $\Delta=\{[1,2,0],[0,1,0],[0,0,1]\}$.
\end{enumerate}
Then the ESPM problem for programs in $\calF(\Delta)$ is NP-complete.
\end{lemma}
Proof: (1)-(2) In each case, if $P\in \calF(\Delta)$, then supported
models of $P$ and answer sets of $P$ coincide (by the properties (S3)
and (S4)).
Thus, the result follows from Lemma \ref{lem1}(1-2), respectively.

\smallskip
\noindent
(3)-(6) In each of these cases, the ESPM problem is equivalent to the
problem of the existence of a model of the program (by Lemma 
\ref{lem:supp1}). Since CNF theories considered in Lemma \ref{lem2}(1-4)
can be written as programs in the classes considered in (3-6), 
respectively, and the rewriting does not affect models, the assertions 
follow by Lemma \ref{lem2}.

\smallskip
\noindent
(7) Let $\vph$ be a CNF formula whose every clause has three literals,
and let $X$ be the set of atoms occurring in $\vph$. For each atom 
$z\in X$ we introduce a fresh atom $z'$. Moreover, for each clause
$c$, we introduce a fresh atom $\hat{c}$. We define a program $P_\vph$
as follows. We include in $P_\vph$ clauses of the form $x\leftarrow x$,
$x'\leftarrow x'$, and $\leftarrow x,x'$, for every $x\in X$. Next, for
each clause $c$ and each atom $x$ that occurs non-negated in $c$, we 
include in $P_\vph$ the clause $\hat{c} \leftarrow x$, and for each atom 
$y$ that occurs negated in $c$, we include in $P_\vph$ the clause 
$\hat{c} \leftarrow y'$. Finally, we include in $P_\vph$ all clauses
$\leftarrow \n \hat{c}$, where $c$ is a clause. Clearly, $P_\vph$
belongs to $\calF(\{[1,1,0],[0,2,0],[0,0,1]\})$. Let $M$ be a model of
$\vph$. We define $U=M\cup\{x'\st x\in X\setminus M\}\cup\{\hat{c}\st
\mbox{$c$ is a clause of $\vph$}\}$. One can check that $U$ is a 
supported model of $P_\vph$. Conversely, let $U$ be a supported model
of $P_\vph$. Let $M=U\cap X$. Let $c$ be a clause in $\vph$. It follows
that $\hat{c}\in U$. Thus, either $c$ has a non-negated atom $x$, such 
that $x\in U$, or a negated atom $y$, such that $y'\in U$. In the first
case, $x\in M$, and so, $M$ is a model of $c$. In the second case, $y
\notin M$ (due to the constraint $\leftarrow y,y'$) and so, $M$ is a 
model of $c$, too.  Thus, $M$ is a model of $\vph$. It follows that the
satisfiability problem for 3CNF formulas can be reduced to the ESPM
problem for $\calF(\{[1,1,0],[0,2,0],[0,0,1]\})$, which implies the
assertion.

\smallskip
\noindent
(8) We can use a similar reduction to the one we employed in (7). We need
one more fresh atom, say $f$ and we modify $P_\vph$ by replacing each 
constraint $\leftarrow x,x'$ with $f\leftarrow x,x'$, and by adding 
the constraint $\leftarrow f$. By this change, we ``simulate'' the 
constraints $\leftarrow x,x'$ within the syntactic confines of the class
$\calF(\{[1,2,0],[0,1,0],[0,0,1]\})$ and so, do not affect the existence
of supported models. \hfill$\Box$

We are now ready to establish the main result of this section.

\begin{theorem}
\label{thm:main:supp}
Let $\Delta\subseteq\calA$. The ESPM problem for $\calF(\Delta)$ is
in P if $\Delta$ satisfies at least one of the following
conditions:
\begin{enumerate}
\item $\Delta\preceq\{[1,0,0],[0,\infty,\infty]\}$
\item $\Delta\preceq\{[\infty,\infty,0],[0,0,0]\}$
\item $\Delta\preceq\{[\infty,1,0],[0,1,0]\}$
\item $\Delta\preceq\{[1,\infty,0],[0,\infty,0]\}$
\item $\Delta\preceq\{[2,0,0],[1,1,0],[0,2,0]\}$
\item $\Delta\preceq\{[1,\infty,0],[0,0,\infty]\}$
\item $\Delta\preceq\{[1,1,0],[0,1,\infty]\}$.
\end{enumerate}
Otherwise, the ESPM problem for $\calF(\Delta)$ is NP-complete.
\end{theorem}
Proof: If $\Delta$ satisfies any of the conditions listed above, the assertion
follows by Lemma \ref{lem:supp:lower}. Thus, let us assume otherwise. We will
show that the ESPM problem for $\calF(\Delta)$ is NP-complete. By the property
(S6), it suffices to show the NP-hardness.

If $[1,0,1]\preceq\Delta$, then the NP-hardness of the ESPM problem for
$\calF(\Delta)$ follows by Lemma \ref{lem:supp:lower}. Thus, let 
$[1,0,1]\not\preceq\Delta$. Under this assumption, we have
$\Delta\preceq\{[\infty,\infty,0],$ $[0,\infty,\infty]\}$. 

We will first assume that $[0,0,1]\preceq\Delta$. If also $[2,0,0]\preceq
\Delta$, Lemma \ref{lem:supp:lower}(2) applies and implies the assertion.
Thus, we can assume that $[2,0,0]\not\preceq\Delta$. Consequently,
$\Delta\preceq\{[1,\infty,0],[0,\infty,\infty]\}$. Since $\Delta\not\preceq
\{[1,\infty,0],[0,0,\infty]\}$, $[0,1,0]\preceq\Delta$. If also $[1,2,0]
\preceq\Delta$, then Lemma \ref{lem:supp:lower}(8) applies. 
Otherwise, $\Delta\preceq\{[1,1,0],[0,\infty,\infty]\}$. Since $\Delta\not
\preceq\{[1,1,0],[0,1,\infty]\}$, $[0,2,0]\preceq\Delta$. Similarly,
since $\Delta\not\preceq\{[1,0,0],[0,\infty,\infty]\}$, $[1,1,0]\preceq\Delta$. 
Thus, Lemma \ref{lem:supp:lower}(7) applies.

We can therefore assume that $[0,0,1]\not\preceq\Delta$. It follows that
$\Delta\preceq\{[\infty,\infty,0], [0,\infty,0]\}$. Since $\Delta\not
\preceq\{[\infty,\infty,0], [0,0,0]\}$, $[0,1,0]\preceq\Delta$. 

Let us assume that $[3,0,0]\preceq\Delta$. If also $[0,2,0]\preceq\Delta$,
then Lemma \ref{lem:supp:lower}(3) applies. Otherwise, we have $\Delta
\preceq\{[\infty,\infty,0], [0,1,0]\}$. Since $\Delta\not\preceq\{[\infty,1,0],
[0,1,0]\}$, $[1,2,0]\preceq\Delta$ and so, Lemma \ref{lem:supp:lower}(5) 
applies. That completes the reasoning for the case when $[3,0,0]\preceq\Delta$.
Therefore, from now on we will assume that $\Delta\preceq\{[2,\infty,0],
[0,\infty,0]\}$.

Since $\Delta\not\preceq\{[1,\infty,0], [0,\infty,0]\}$, $[2,0,0]\preceq
\Delta$. If $[0,3,0]\preceq\Delta$, Lemma \ref{lem:supp:lower}(6) applies. 
Otherwise, $\Delta\preceq\{[2,\infty,0], [0,2,0]\}$. Since $\Delta\not
\preceq\{[2,0,0],[1,1,0],[0,2,0]\}$, we have $[1,2,0]\preceq\Delta$ or
$[2,1,0]\preceq\Delta$. In the former case, Lemma \ref{lem:supp:lower}(5)
applies. Thus, let us assume that $[2,1,0]\preceq\Delta$. If $[0,2,0]\preceq
\Delta$, then Lemma \ref{lem:supp:lower}(4) applies. Otherwise, we have
$\Delta\preceq\{[2,\infty,0],[0,1,0]\}$. Since $\Delta\not\preceq
\{[\infty,1,0], [0,1,0]\}$, it follows that $[1,2,0]\preceq\Delta$, the case
that we handled above. \hfill$\Box$

Next, we establish the complexity of the problem to decide
whether a program $P$ credulously entails under the supported-model
semantics the literal $\neg a$, where $a$ is an atom.

\begin{theorem}
\label{thm:supp:cr}
Let $\Delta\subseteq\calA$. If
\begin{enumerate}
\item $\Delta\preceq\{[1,0,0],[0,\infty,\infty]\}$, or
\item $\Delta\preceq\{[\infty,1,0],[0,1,0]\}$, or
\item $\Delta\preceq\{[1,\infty,0],[0,\infty,0]\}$, or
\item $\Delta\preceq\{[2,0,0],[1,1,0],[0,2,0]\}$, or
\item $\Delta\preceq\{[1,1,0],[0,1,\infty]\}$,
\end{enumerate}
then the problem to decide, given an atom $a$ and a program $P\in
\calF(\Delta)$, whether $P$ has a supported model that does not contain 
$a$ is in $\mathrm{P}$. Otherwise, the problem is $\mathrm{NP}$-complete.
\end{theorem}
Proof: The problem is clearly in NP.
Let $\Delta$ satisfy one of the conditions listed above. Let $P
\in\calF(\Delta)$ and let $a$ be an atom. We define $P'=P\cup\{\leftarrow a\}$.
Clearly, $P$ has a supported model that does not contain $a$ if and only if
 $P'$ has a supported model. Since $P'\preceq\calF(\Delta)$, the assertion 
follows by Lemma \ref{lem:supp:upper}. 

Thus, from now on, we assume that $\Delta$ does not satisfy any of the
conditions (1)-(5). Let us start by considering the case $\Delta\preceq
\{[\infty,\infty,0],[0,0,0]\}$. Since $\Delta\not\preceq\{[1,\infty,0],$
$[0,\infty,0]\}$, it follows that $[2,0,0]\preceq\Delta$. Moreover, since 
$\Delta
\not\preceq\{[\infty,1,0], [0,1,0]\}$, $[1,2,0]\preceq\Delta$. Let $P\in
\calF(\{[2,0,0],[1,2,0], [0,1,0]\})$. We select a fresh atom $a$ and 
define $P'=\sov{P}\cup\{a \leftarrow \bd(r)\st r\in\dov{P}\}$. Clearly, 
$P'\in \calF(\Delta)$. Moreover, $P$ has a supported model if and only 
if $P'$ has a supported model that does not contain $a$.
Thus, the claim follows by Lemma \ref{lem:supp:lower}(5).

Next, let us assume that $\Delta\preceq\{[1,\infty,0],[0,0,
\infty]\}$. Since $\Delta\not\preceq\{[1,\infty,0],[0,\infty,0]\}$,
$[0,0,1]\preceq\Delta$.
Similarly, since $\Delta\not\preceq\{[1,1,0],[0,1,\infty]\}$, 
it follows that $[1,2,0]\preceq\Delta$. Let $P\in\calF(\{[1,2,0],[0,1,0],
[0,0,1]\})$. We select a fresh atom $a$ and define $P'=\sov{P}\cup\{a
\leftarrow b\st \leftarrow b \in\dov{P}\}\cup\{\leftarrow \n b\st 
\leftarrow \n b \in \dov{P}\}$. Clearly, $P'\in \calF(\Delta)$.
Moreover, $P$ has a supported model if and only if $P'$ has a supported 
model that does not contain $a$. Thus, the claim follows by Lemma 
\ref{lem:supp:lower}(8).

If $\Delta\not\preceq\{[\infty,\infty,0],[0,0,0]\}$ and $\Delta\not
\preceq\{[1,\infty,0],[0,0,\infty]\}$ then, by Theorem \ref{thm:main:supp},
the ESPM problem for the class $\calF(\Delta)$ is NP-complete. Let
$P\in\calF(\Delta)$ and let $a$ be a fresh atom. Clearly, $P$ has a supported
model if and only if $P$ has a supported model that does not contain $a$.
Thus the assertion follows. \hfill$\Box$

Finally, we settle the complexity of the problem to decide whether a
program $P$ skeptically entails under the supported model semantics
the literal $\neg a$, where $a$ is an atom.
 
\begin{theorem}
\label{thm:supp:sk}
Let $\Delta\subseteq\calA$. If
\begin{enumerate}
\item $\Delta\preceq\{[1,0,0],[0,\infty,\infty]\}$, or
\item $\Delta\preceq\{[1,\infty,0],[0,0,\infty]\}$, or
\item $\Delta\preceq\{[1,1,0],[0,1,\infty]\}$, or
\item $\Delta\preceq\{[1,1,0],[0,\infty,0]\}$,
\end{enumerate}
then the problem to decide, given an atom $a$ and a program $P\in
\calF(\Delta)$, whether every supported model of $P$ satisfies $\neg a$
is in $\mathrm{P}$. Otherwise, the problem is $\mathrm{coNP}$-complete.
\end{theorem}
Proof: We prove the assertion by establishing the complexity of the
complementary problem to decide, given an atom $a$ and a program $P\in
\calF(\Delta)$, whether $P$ has a supported model that contains $a$.
We will denote this problem by $\calC$.

Let $\Delta$ satisfy any of the conditions (1)-(3), $P\in\calF(\Delta)$
and $a$ be an atom. We define $P'=P\cup\{\leftarrow\n a\}$. Clearly,
$P$ has a supported model containing $a$ if and only if $P'$ has a 
supported model. Since $P'\in\calF(\Delta \cup \{[0,0,1]\})$ and
$\Delta \cup \{[0,0,1]\}$ falls under the same condition as $\Delta$ 
does, the latter property can be verified in polynomial time (cf. Lemma
\ref{lem:supp:upper}). Thus, the problem $\calC$ can be decided in
polynomial time.

Thus, let us suppose that $\Delta\preceq\{[1,1,0],[0,\infty,0]\}$.
Let $P\in\calF(\{[1,1,0]\})$. Since $P$ is a Horn program, then the least 
model of $P$, $LM(P)$, is a supported model of $P$. One can show that for
every atom $b$ such that there is a sequence of rules $b_0\leftarrow b_1$; 
$b_1\leftarrow b_2;\ldots; b_{k-1}\leftarrow b_k$ in $P$, where $b_0=b_k=b$
and $k\geq 1$, the set $LM(P\cup\{b\})$ is a supported model of $P$.
Moreover, one can show that every supported model of $P$ is the union of 
some of the sets $LM(P)$ and $LM(P\cup\{b\})$, where $b$ is self-supported
(in the sense that it participates in a cycle of rules, as described above).
It follows that $P\in\calF(\{[1,1,0],[0,\infty,0]\})$ has a supported
model containing $a$ if and only if one of the sets $LM(\sov{P}\cup\{b\})$,
where $b$ is an atom, is a supported model of $P$ that contains $a$. The
latter property can be verified in polynomial time. Thus the problem $\calC$
is in P for the class of programs $\calF(\Delta)$, where
$\Delta\preceq \{[1,1,0],[0,\infty,0]\})$.
 
To complete the proof, let us assume that $\Delta$ does not satisfy
any of the conditions (1)-(4). If $[2,0,0]\preceq\Delta$ then the 
NP-completeness of the problem $\calC$ follows by the result of Cadoli
and Lenzerini \citeyear{cl94}, in the same way as in the proof of Theorem
\ref{thm:cr} (we note that, by the properties (S3) and (S4), for programs 
in the class $\calF(\{[2,0,0]\})$ supported models and answer sets coincide). 

Thus, let us assume that $[2,0,0]\not\preceq\Delta$, that is, $\Delta
\preceq\{[1,\infty,\infty],[0,\infty,\infty]\}$.
For programs in $\calF(\{[1,0,1]\})$, the problem of the existence of an 
answer set that contains a given atom $a$ is NP-complete (it follows,
in particular, from Theorem \ref{thm:sk}). Since for programs in that
class answer sets and supported models coincide (by the properties (S3)
and (S4)), the problem of the existence of a supported
model of $P\in\calF(\{[1,0,1]\})$ that contains a given atom $a$ is 
NP-complete, too. Thus, the assertion holds for every class $\Delta$ such 
that $[1,0,1]\preceq\Delta$. It follows that it is enough to prove the
theorem under the assumption that $\Delta\preceq\{[1,\infty,0],[0,\infty,
\infty]\}$.

Since $\Delta\not\preceq\{[1,\infty,0],[0,0,\infty]\}$, $[0,1,0]\preceq
\Delta$. Let us assume that $[1,2,0]\preceq\Delta$. Let $P\in
\calF(\{[1,2,0],[0,1,0],[0,0,1]\})$. Let $a_1,\ldots a_k$ be all the
atoms such that the constraint $\leftarrow \n a_i$ belongs to $P$. 
We introduce fresh atoms $b_1,\ldots,b_k$ and define $P'$ to be obtained
by replacing all constraints $\leftarrow \n a_i$ with rules $b_1\leftarrow
a_1$ and $b_i\leftarrow b_{i-1}, a_i$. Clearly, $P'\in\calF(\{[1,2,0],
[0,1,0]\})$. One can check that $P$ has a supported model if and only if  
$P'$ has a supported model containing $b_k$. Thus, in this case, the
problem is NP-complete. It follows that $\Delta\preceq\{[1,1,0],[0,\infty,
\infty]\}$. Since $\Delta\not\preceq\{[1,0,0],[0,\infty,\infty]\}$, 
$[1,1,0]\preceq\Delta$. Next, since $\Delta\not\preceq\{[1,1,0],[0,1,
\infty]\}$, $[0,2,0]\preceq\Delta$. Finally, since $\Delta\not\preceq
\{[1,1,0],[0,\infty,0]\}$, $[0,0,1]\preceq\Delta$. Thus, the ESPM problem
for $\calF(\Delta)$ is NP-complete (Lemma \ref{lem:supp:lower}(7)). 
Let $P\in\calF(\Delta)$
and let $a$ be a fresh atom. We define $P'=P\cup\{a\leftarrow a\}$. Clearly,
$P'$ has a supported model containing $a$ if and only if $P$ has a supported
model. Thus, the NP-completeness of the problem $\calC$ follows. 
\hfill$\Box$

\section{Discussion}

In the paper, we studied classes of programs defined in terms of
``legal'' arities of rules. Specifically, we focused  on classes of
programs of the form $\calF(\Delta)$, where $\Delta\subseteq\calA$. We
proved that each such class has a finite representation and, for
each finite set $\Delta$, we determined the complexity of reasoning
tasks for programs from $\calF(\Delta)$ under the answer-set semantics. 
We also considered briefly a related family of classes of programs, 
namely those of the form $\calG(\Delta)$, where $\Delta\subseteq\calA$ 
consists of arities only, and obtained similar results for them. 
Our results can be regarded as \emph{trichotomy} 
results as, in each case, the complexity is given by one of three 
complexity classes (P, NP-complete, and $\Sigma_2^P$-complete; or 
P, coNP-complete, and $\Pi_2^P$-complete, depending on the type of the 
reasoning task). We also presented similar results for the reasoning tasks
under the supported-model semantics. However, in that case, we obtain 
\emph{dichotomy} results --- the complexity is given either by the class 
P or NP-complete (or, either by the class P or coNP-complete, depending on
the task). 

As we noted, our trichotomy and dichotomy results have some similarity 
to the dichotomy result by Schaefer, and its corollaries for other logic 
formalisms: the abductive reasoning \cite{NordhZ08}, reasoning with 
minimal models \cite{Cadoli92} and, reasoning in default logic 
\cite{ChapdelaineHS07}. 
The classes of theories and formulas considered in those papers are
defined in terms of Boolean relations that are allowed in the language 
\cite{Schaefer78,BulatovJK05,CreignouKS01}. That definition schema
satisfies the dichotomy property: for every class of formulas definable
in that schema, the satisfiability problem is in P, or is NP-complete.
The monotonicity of the propositional logic (the set of models of the 
conjunction of two formulas is the intersection of the sets of models 
of the conjuncts) is a fundamental property required by that result.
Since logic programs with the answer-set semantics do not satisfy the
monotonicity property, it is unclear how to extend that formalism
the approach originated by Schaefer. Thus, we based our approach on a 
different definition schema developed specifically for programs, and 
related to the ``complexity'' of rules as measured by the numbers of 
atoms in the head, and positive and negative literals in the body. 

It turns out though, that some classes of programs/theories appear 
prominently in both settings (for instance: Horn programs and Horn 
theories; positive
programs with no more than two literals per rule and 2CNF theories). It
is then  an interesting problem whether a result based on the 
classification in terms of types of Boolean relations can be obtained 
for disjunctive logic programs. One possibility might be to consider 
a more general setting of answer-set programs in the language of 
propositional logic under the semantics of equilibrium models 
\cite{fl05}.

\section*{Acknowledgments}
\vspace*{-0.1in}
This paper is an extended version of the paper presented at the 10th
International Conference on Logic Programming and Nonmonotonic Reasoning
\cite{tru09}. The work was partially supported by the NSF grant IIS-0913459.
The author gratefully acknowledges several helpful comments from the
anonymous referees.


\end{document}